\documentclass[referee]{raa}
\usepackage{graphicx,times}
\usepackage{natbib}
\usepackage{amssymb,amsmath}
\bibpunct{(}{)}{;}{a}{}{,}
\usepackage[a4paper=true,dvipdfm=true,pagebackref=true]{hyperref}
\hypersetup{pdftitle = The title of my PDF, pdfauthor = My name, pdfsubject= The subject, pdfkeywords = keyword1 keyword2 keyword3}
\hypersetup{colorlinks = true, linkcolor = green, anchorcolor = red, citecolor = blue, filecolor = red, pagecolor = red, urlcolor = red}

\begin{document}

\title{Modeling the evolution and distribution of the frequency's
second derivative and the braking index of pulsar spin}

\volnopage{ {\bf 2014} Vol.\ {\bf X} No. {\bf XX}, 000--000} \setcounter{page}{1}

\author{Yi Xie\inst{1,2}, Shuang-Nan Zhang\inst{1,3,4}, and Jin-Yuan Liao\inst{2,3}
   }
\institute{National Astronomical Observatories, Chinese Academy of Sciences, Beijing 100012, China\\
\and
University of Chinese Academy of Sciences, Beijing 100049, China\\
\and
Key Laboratory of Particle Astrophysics, Institute of High Energy Physics, Chinese Academy of Sciences, Beijing 100049, China; {\it zhangsn@ihep.ac.cn}\\
\and
Physics Department, University of Alabama in Huntsville, Huntsville, AL 35899, USA\\ }

\abstract{We model the evolution of the spin frequency's second derivative $\ddot\nu$ and the braking index $n$ of radio pulsars with simulations
within the phenomenological model of their surface magnetic field evolution, which contains a long-term power-law decay modulated by short-term
oscillations. For the pulsar PSR B0329+54, a model with three oscillation components can reproduce its $\ddot\nu$ variation. We show that the ``averaged'' $n$ is different from the instantaneous $n$, and its oscillation magnitude decreases abruptly as the time span increases, due to the ``averaging'' effect. The simulated timing residuals agree with the main features of the reported data. Our model predicts that the averaged $\ddot\nu$ of PSR B0329+54 will start to decrease rapidly with newer data beyond those used in Hobbs et al.. We further perform Monte Carlo simulations for the distribution of the reported data in $|\ddot\nu|$ and $|n|$ versus characteristic age $\tau_{\rm c}$ diagrams. It is found that the  magnetic field oscillation model with decay index $\alpha=0$ can reproduce the distributions quite well. Compared with magnetic field decay due to the ambipolar diffusion ($\alpha=0.5$) and the Hall cascade ($\alpha=1.0$), the model with no long term decay ($\alpha=0$) is clearly preferred for old pulsars by the p-values of the two-dimensional Kolmogorov-Smirnov test.
}

   \authorrunning{Y. Xie, S.N. Zhang, and J.Y. Liao. }            
   \titlerunning{Modeling the frequency's second derivative of pulsar spin}  
   \maketitle

%
\section{Introduction}           
\label{sect:intro}

The spin-down of radio pulsars is caused by emitting electromagnetic
radiation and by accelerating particle winds. Traditionally, the
evolution of their rotation frequencies $\nu$ may be described by
the braking law
\begin{equation}\label{braking law}
\dot\nu =-K \nu^n,
\end{equation}
where $n$ is the braking index, and $K$ is a positive constant that
depends on the magnetic dipole moment and the moment of inertia of
the neutron star (NS). By differentiating Equation (\ref{braking law}),
one can obtain $n$ in terms of several observables, $ n=\ddot\nu
\nu/\dot\nu^2$. For the standard vacuum magnetic dipole radiation
model with a constant magnetic field (i.e. $\dot K=0$), $n=3$
(Manchester \& Taylor 1977). Thus the frequency's second derivative
can be simply expressed as
\begin{equation}\label{ddotnu}
\ddot\nu=3\dot\nu^2/\nu.
\end{equation}
The model predicts $\ddot\nu>0$ and $|\ddot\nu|$ should be very
small.

However, unexpectedly large values of $\ddot\nu$ were measured for
several dozen pulsars thirty years ago (Gullahorn \& Rankin 1978;
Helfand et al. 1980; Manchester \& Taylor 1977), and many of those
pulsars surprisingly showed $\ddot\nu<0$. Some authors suggested
that the observed values of $\ddot\nu$ could result from a
noise-type fluctuation in the pulsar period (Helfand et al. 1980;
Cordes 1980; Cordes \& Helfand 1980). Based on the timing data of
PSR B0329+54, Demia${\rm \acute{n}}$ski \& Pr${\rm
\acute{o}}$szy${\rm \acute{n}}$ski (1979) further proposed that a
distant planet would influence $\ddot\nu$, and the quasi-sinusoidal
modulation in timing residuals might be caused by changes in pulse
shape, precession of a magnetic dipole axis, or an orbiting planet.
Baykal et al. (1999) investigated the stability of $\ddot\nu$ for pulsars PSR
B0823+26, PSR B1706-16, PSR B1749-28 and PSR B2021+51 using their
time-of-arrival (TOA) data extending to more than three decades. This analysis
confirmed that the anomalous $\ddot\nu$ terms of these sources
arise from red noise (timing residuals with low frequency
structure), which may originate from external torques applied by the
magnetosphere of a pulsar.

The relationship between the low frequency structure in timing
residuals and the fluctuations in pulsar spin parameters ($\nu$,
$\dot\nu$, and $\ddot\nu$) is very interesting and important. We
call both the residuals and the fluctuations the ``timing noise''
in the present work, since we will infer that they have the same
origin. Timing noise for some pulsars has been studied for over
four decades (e.g. Boynton et al. 1972; Groth 1975; Jones 1982;
Cordes \& Downs 1985; D'Alessandro et al. 1995; Kaspi, Chakrabarty
\& Steinberger 1999; Chukwude 2003; Livingstone et al. 2005; Shannon
\& Cordes 2010; Liu et al. 2011; Coles et al. 2011; Jones 2012).
However, the origins of the timing noise are still controversial and
there is still unmodelled physics to be understood. Boynton et al.
(1972) suggested that the timing noise might arise from ``random
walk'' processes. The random walk in $\nu$ may be produced by small
scale internal superfluid vortex unpinning (Alpar, Nandkumar \&
Pines 1986; Cheng 1987a), or short time ($t\sim 10~{\rm ms}$ for the
Crab pulsar) fluctuations in the size of the outer magnetosphere gap
(Cheng 1987b). Stairs, Lyne \& Schemar (2000) reported long
time-scale, highly periodic and correlated variations in the pulse
shape and the slow-down rate of the pulsar PSR B1828-11, which have
generally been considered as evidence of free precession. The
possibilities were also proposed that the quasi-periodic modulations
in timing residuals could be caused by an orbiting asteroid belt
(Cordes \& Shannon 2008) or a fossil accretion disk (Qiao et al.
2003).

Recently, Hobbs et al. (2010, hereafter H2010) carried out the
most extensive study so far of long term timing irregularities of 366
pulsars. Besides ruling out some timing noise models in terms of
observational imperfections, random walks, and planetary companions,
some of their main conclusions were: (1) timing noise is widespread
in pulsars and is inversely correlated with pulsar characteristic
age $\tau_{\rm c}$; (2) significant periodicities are seen in the
timing noise of a few pulsars, but quasi-periodic features are
widely observed; (3) the structures seen in the timing noise vary
with data spans, i.e., more quasi-periodic features are seen for a longer
data span and the magnitude of $|\ddot{\nu}|$ for a shorter data span
is much larger than that caused by the magnetic braking of the NS; and (4) the numbers of negative and positive $\ddot{\nu}$ are
almost equal in the sample, i.e. $N_{\rm p}\thickapprox
N_{\rm n}$. Lyne et al. (2010) showed credible evidence that
timing noise and $\dot\nu$ are correlated with changes in the pulse
shapes, and are therefore linked to and caused by changes in the
pulsar's magnetosphere.

Blandford \& Romani (1988) re-formulated the braking law of a pulsar
as $\dot\nu=-K(t)\nu^3$, which means that the standard magnetic
dipole radiation is still responsible for the instantaneous
spin-down of a pulsar, and $\ddot\nu\nu/\dot\nu^2\neq3$ does not
indicate deviation from the dipole radiation model, but only means
that $K(t)$ is time dependent. Considering the magnetospheric origin
of timing noise as inferred by Lyne et al. (2010), we assume that
magnetic field evolution is responsible for the variation of $K(t)$,
i.e. $K=A B(t)^2$, in which $A=\frac{8\pi^2R^6\sin\theta^2}{3c^3I}$
is a constant, and $R~(\simeq10^6~{\rm cm})$, $I~(\simeq10^{45}~{\rm
g~cm^2})$, and $\theta~(\simeq\pi/2)$ are the radius, moment of
inertia, and angle of magnetic inclination for the NS,
respectively. We can rewrite Equation (\ref{ddotnu}) as
\begin{equation}\label{ddotnu-2}
\ddot\nu=3\dot\nu^2/\nu+2\dot\nu\dot B/B.
\end{equation}
Since the numbers of negative and positive $\ddot{\nu}$ are almost
equal, it should be the case that $B$ quasi-symmetrically
oscillates, and usually $|2\dot\nu\dot B/B|\gg3\dot\nu^2/\nu$.
In addition, it can be noticed that pulsars with $\tau_{\rm
c}\lesssim10^5~{\rm yr}$ always have $\ddot\nu>3\dot\nu^2/\nu$
(H2010); a reasonable interpretation is that their magnetic field
decays (i.e. $\dot B<0$) dominate the field evolution for these
``young'' pulsars.

Therefore, Zhang \& Xie (2012a, hereafter Paper I) constructed a
phenomenological model for the dipole magnetic field evolution of
pulsars with a long-term decay modulated by short-term oscillations,
\begin{equation}\label{B evolution}
B(t)=B_d(t)(1+\sum k_i\sin(\phi_i+2\pi\frac{t}{T_i})),
\end{equation}
where $t$ is the pulsar's age, and $k_i$, $\phi_i$ and $T_i$ are the
amplitude, phase and period of the $i$-th oscillating component,
respectively. $B_d(t)=B_0(t/t_0)^{-\alpha}$, in which $B_0$ is the
field strength at the age $t_0$, the index $\alpha=0$ means the field has no long-term decay, and it was found that $\alpha\gtrsim0.5$ for young pulsars with $\tau_{\rm c}<10^6$ yr (see Paper I for details). Substituting Equation (\ref{B evolution})
into Equation (\ref{braking law}), we get the differential equation
describing the the spin frequency evolution of a pulsar as follows
\begin{equation}\label{braking law2}
\dot\nu \nu^{-3}=-A B(t)^2.
\end{equation}

In paper I, we showed that the distribution of $\ddot\nu$ and the
inverse correlation of $\ddot\nu$ versus $\tau_c$ could be
explained well with analytic formulae derived from the phenomenological
model. In Zhang \& Xie (2012b, hereafter Paper II), we also derived
an analytical expression for the braking index ($n$) and pointed out
that the instantaneous value of $n$ of a pulsar is different from
the ``averaged'' $n$ obtained from the traditional phase-fitting
method over a certain time span. However, this ``averaging'' effect
was not included in our previous analytical studies; this work is
focused on addressing this effect.

This paper is organized as follows. In Section 2, we show that the
timescales of magnetic field oscillations are tightly connected to
the $\ddot\nu$ evolution and the quasi-periodic oscillations
appearing in the timing residuals, and the reported data of PSR
B0329+54 are fitted. In Section 3, we perform Monte Carlo
simulations on the pulsar distribution in the $\ddot\nu-\tau_{\rm
c}$ and $n-\tau_{\rm c}$ diagram. Our results are summarized and
discussed in Section 4.

\section{Modeling The $\ddot\nu$ and $n$ Evolution and Timing Residuals of Pulsar PSR B0329+54}
\label{sect:Obs}

PSR B0329+54 is a bright (e.g. 1500 mJy at 400
MHz\footnote{http://www.atnf.csiro.au/people/pulsar/psrcat/}),
$0.71$ s pulsar that had been suspected of possessing planetary-mass
companions (Demia${\rm \acute{n}}$ski \& Pr${\rm
\acute{o}}$szy${\rm \acute{n}}$ski 1979; Bailes, Lyne, \& Shemar
1993; Shabanova 1995). However the suspected companions have not
been confirmed and their existence is currently considered doubtful (Cordes \&
Downs 1985; Konacki et al. 1999; H2010). Konacki et al. (1999)
suggested that the observed ephemeral periodicities in the timing
residuals for PSR B0329+54 are intrinsic to this NS. H2010
believed that the timing residual has a form that is similar to other
pulsars in their sample. They plotted $|\ddot\nu|$ obtained from the
B0329+54 data sets with various time spans (see Figure 12 in their
paper). For data spanning $\sim 10$ yr, they measured a large and
significant $\ddot\nu$, and found that the timing residuals take the
form of a cubic polynomial. However, no cubic term was found for
data spanning more than $\sim 25$ yr, and $|\ddot\nu|$ became
significantly smaller. The reported periods of the timing
residuals for PSR B0329+54 are $1100~{\rm days}$, $2370~{\rm days}$,
and/or $16.8~{\rm years}$ (Demia${\rm \acute{n}}$ski \& Pr${\rm
\acute{o}}$szy${\rm \acute{n}}$ski 1979; Bailes, Lyne, \& Shemar
1993; Shabanova 1995).

In order to model the $\ddot\nu$ evolution for pulsar PSR B0329+54, we
first obtain $\nu(t)$ by integrating the spin-down law described
by Equations (\ref{B evolution}) and (\ref{braking law2}), and then the phase
\begin{equation}\label{phase integrate}
\Phi(t)=\int_{t_0}^{t}\nu(t'){\rm d}t'.
\end{equation}
Finally, these observable quantities, $\nu$, $\dot{\nu}$ and
$\ddot{\nu}$ can be obtained by fitting the phases to the third
order of its Taylor expansion over a time span $T_{\rm s}$,
\begin{equation}\label{phase}
\Phi(t_i) =\Phi_0 + \nu (t_i-t_0) + \frac{1}{2}\dot \nu (t_i-t_0)^2
+ \frac{1}{6}\ddot\nu (t_i-t_0)^3.
\end{equation}
We thus get $\nu$, $\dot{\nu}$ and $\ddot{\nu}$ for $T_{\rm s}$ from
fitting to Equation~(\ref{phase}), with a certain time interval of
phases $\Delta T_{\rm int}=10^6~{\rm s}$ (interval between each TOA,
i.e. $\Delta T_{\rm int}=t_{i+1}-t_{i}$).

We adopt a goodness of fit parameter to show how well the
model matches the data, i.e. $\chi^2=\sum\chi_{i}^2=\sum\frac{(\ddot\nu_{i \rm M}-\ddot\nu_{i \rm D})^2}{\sigma_{i}^2}$, where the subscripts ${\rm M}$ and ${\rm D}$ refer to the model results and the reported data, and $\sigma_i$ are the uncertainties in the reported data. In order to minimize $\chi^2$, we adopt the Simulated Annealing Algorithm (SAA) to reach a fast convergence and avoid being trapped in a local minimum, and we use a simulation based on the Markov chain Monte Carlo (MCMC) methods for the fitting to explore the whole parameter space.

\begin{figure}
\centering
\includegraphics[angle=0,scale=0.35]{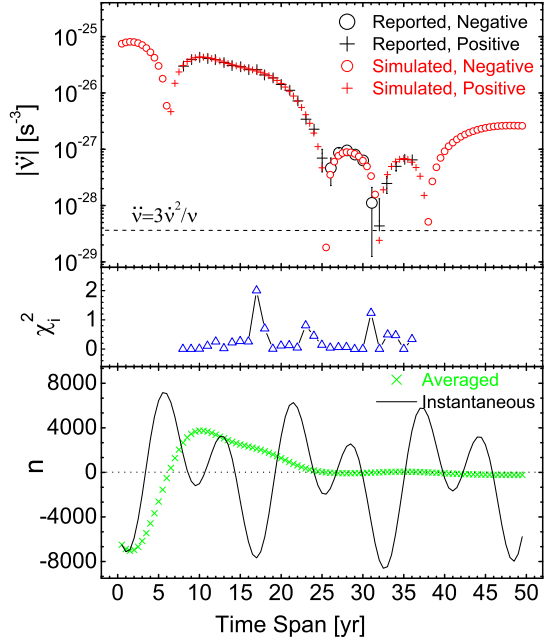}
\caption{$|\ddot\nu|$, $\chi_{i}^2$ and $n$ for PSR B0329+54. Top panel:
reported and fitted $|\ddot\nu|$. The values
reported by H2010 are represented by large cross symbols
($\ddot\nu>0$) and large circles ($\ddot\nu<0$); the best-fitting results are represented by small cross symbols ($\ddot\nu>0$) and
small circles ($\ddot\nu<0$); the horizontal dashed line represents
$\ddot\nu=3\dot\nu^2/\nu$. Middle panel: the goodness of fit parameter $\chi_{i}^2$ for the fit of $|\ddot\nu|$. It is shown that the three-component model fits the reported data quite well. Bottom panel: instantaneous (solid line) and averaged (crosses) values of  $n$. The horizontal dotted line represents $n=0$.} \label{Fig:1}
\end{figure}

In the upper panel of Figure {\ref{Fig:1}}, we show the
reported and the best-fitting (simulated) results of $|\ddot\nu|$ for various $T_{\rm
s}$ for PSR B0329+54; the reported data are read from Figure 12 of
H2010. There are three oscillation components involved in the simulation, and $\alpha=0$ is taken from Equation (\ref{B evolution}). The obtained smallest value of $\chi^2$ is 9.1, with the number of degrees of freedom being $20$, and all the best-fit parameters for the three oscillation components are listed in Table {\ref{Tab:1}}. $\chi_{i}^2$ for each reported data point is also shown in the middle panel; in the bottom panel, we show the corresponding $n$ with the same oscillation parameters obtained above. The braking index $n=\ddot\nu
\nu/\dot\nu^2$ obtained directly from Equation (\ref{braking law2})
is called ``instantaneous'' $n$; similarly, what is obtained by
fitting phase sets to Equation (\ref{phase}) is called ``averaged''
$n$. It can be seen that the averaged $n$ has the same variation
trends as $\ddot\nu$, since $|\Delta\nu/\nu| \sim10^{-6}$
and $|\Delta\dot\nu/\dot\nu| \sim10^{-3}$ are tiny, compared to
$|\Delta\ddot\nu/\ddot\nu| \sim 1$. The magnitude of the first
period of the averaged $n$ is close to the instantaneous one, but it
decays significantly due to the ``averaging'' effect.

Since both the fits with one and two oscillation components
are not very good and are certainly rejected by the $\chi^2$ test (e.g. the smallest $\chi^2$ of the two component simulation is $128$), and $\chi^2$ is not significantly reduced after setting the index $\alpha$ as a free parameter, we thus conclude that three oscillation components are necessary for the fitting the variation in $|\ddot\nu|$.

\begin{figure}
\centering
\includegraphics[angle=0,scale=0.5]{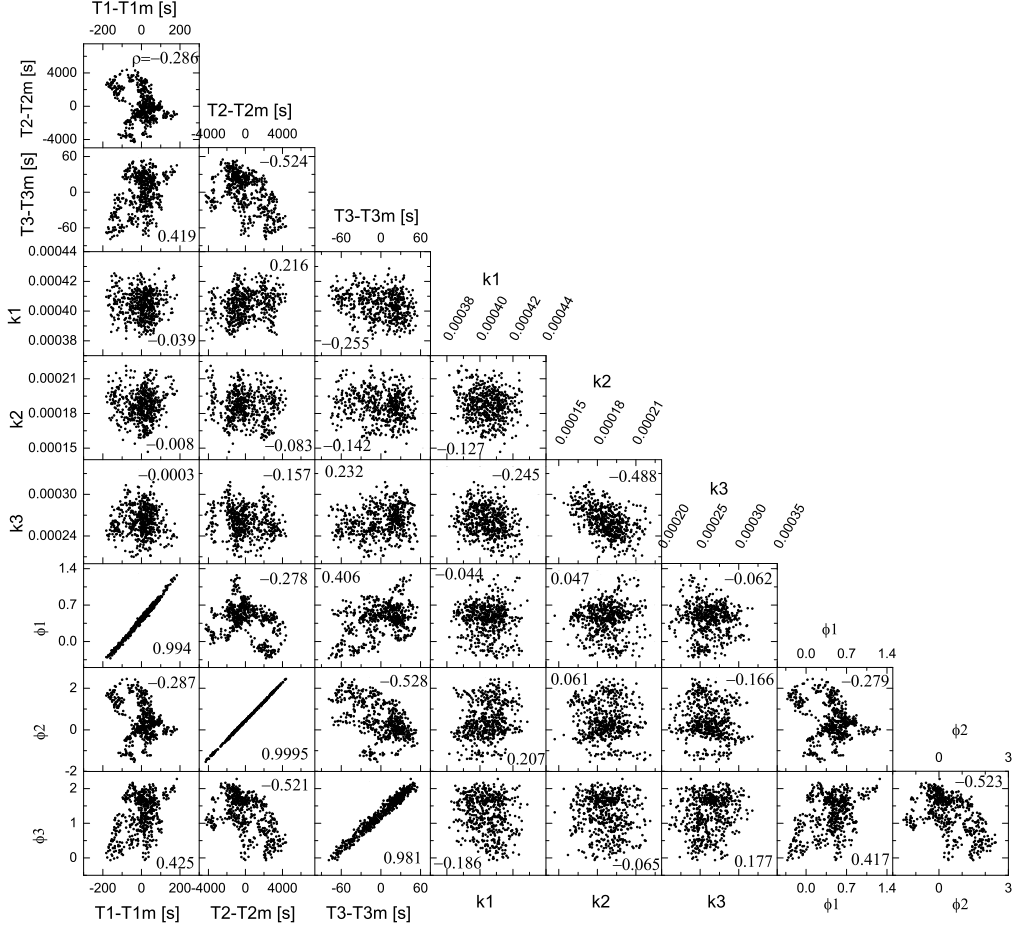}
\caption{Correlation of the nine fitting parameters of $|\ddot\nu|$. Each panel shows the joint posterior probability distribution between a pair of parameters, with correlation coefficient $\rho$ labeled in it. The oscillation periods $T_1$, $T_2$, and $T_3$ are plotted in terms of the differences from their mean values $T_{1\rm m}=5.00482804\times 10^8$~s, $T_{2\rm m}=1.546443564\times 10^9$~s, and $T_{3\rm m}=2.48163290\times 10^8$~s, respectively.} \label{Fig:2}
\end{figure}

We use the Pearson Correlation Coefficient $\rho=\frac{{\rm Covariance}(X,Y)}{\sqrt{{\rm Variance}(X){\rm Variance}(Y)}}$ to measure the covariance between the parameters, where $X$ and $Y$ are the parameters to be tested. We show the joint posterior probability distribution between each pair of parameters in Figure \ref{Fig:2}, with $\rho$ labeled in each panel. For each of the three oscillation components, their phase $\phi$ is completely coupled with their period $T$. All other parameters are well determined independently.

\begin{table}
\bc
\caption[]{Summary of all the
best-fitting parameters. The first row lists the best-fitting values for all the parameters, and the second row lists their $1\sigma$ errors.\label{Tab:1}}
\setlength{\tabcolsep}{4pt}
 \begin{tabular}{lccccccccc}
  \hline\noalign{\smallskip}
Parameters & $T_1$ & $T_2$ & $T_3$ & $k_1$       & $k_2$       & $k_3$       & $\phi_1$ & $\phi_2$ & $\phi_3$\\
~          & (yr)  & (yr)  &(yr)   & ($10^{-4}$) & ($10^{-4}$) & ($10^{-4}$) & (rad)    & (rad)    & (rad)   \\
  \hline\noalign{\smallskip}
Best-fitting  & 15.870206           & 49.03738            & 7.869207            & 4.05 & 1.89 & 2.58 & 0.496 & 0.071 & 1.937 \\
$1\sigma$ error & $5.1\times 10^{-6}$ & $1.6\times 10^{-4}$ & $2.5\times 10^{-6}$ & 0.11 & 0.12 & 0.21 & 0.148 & 0.36 & 0.427  \\
  \noalign{\smallskip}\hline
\end{tabular}
\ec
\end{table}

The timing residuals, after subtraction of the pulsar's $\nu$ and
$\dot\nu$ over $36.5$ years for PSR B0329+54, are also simulated with
exactly the same model parameters used for modeling $\ddot\nu$. In
the simulation, the following steps are taken:

(i) We get the model-predicted TOAs with $\Delta T_{\rm
int}=10^6~{\rm s}$ using Equation (\ref{phase integrate}) over
$36.5~{\rm yr}$, with the same model parameters used for modeling
$\ddot\nu$.

(ii) By fitting the TOA set $\{\Phi(t_i)\}$ to
\begin{equation}\label{phase2}
\Phi(t) =\Phi_0 + \nu_0 (t-t_0) + \frac{1}{2}\dot\nu_0 (t-t_0)^2,
\end{equation}
we get $\Phi_0$, $\nu_0$ and $\dot\nu_0$.

(iii) Then the timing residuals after the subtraction of $\nu$ and
$\dot\nu$ can be obtained by
\begin{equation}\label{residual}
T_{\rm res}(t_i)=\frac{\Phi(t_i)-(\Phi_0 + \nu_0 (t_i-t_0) +
\frac{1}{2}\dot\nu_0 (t_i-t_0)^2)}{\nu_0}.
\end{equation}

\begin{figure}
\centering
\includegraphics[angle=0,scale=0.35]{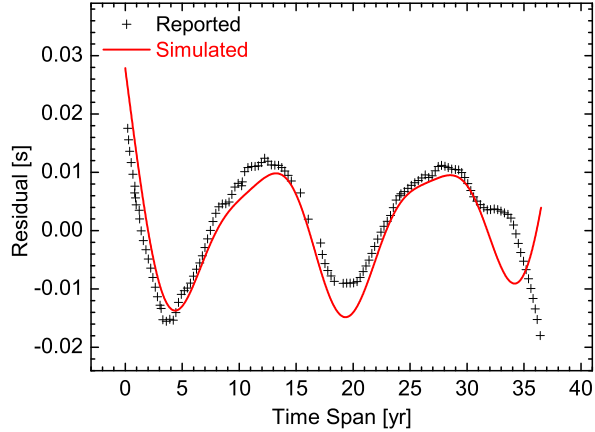}
\caption{Timing residuals of PSR B0329+54. The reported timing residual, after subtraction $\nu$ and $\dot\nu$ of the pulsar over
the $36.5$ years, is represented by crosses; the predicted residuals modeled by three oscillation components are represented by the solid line. The model parameters are identical to those for the $\ddot\nu$ simulation, as shown in Figure {\ref{Fig:1}}. } \label{Fig:3}
\end{figure}

In Figure \ref{Fig:3}, we plot the reported timing residuals (from fig. 3 of H2010) with crosses, and the
simulated result of the model with three oscillation components
with a solid line. Note that the simulated result is not the fit of the model to the reported timing
residual. It is actually the application of the model with the parameters derived from the fitting of $|\ddot\nu|$, i.e. the figure shows a comparison
of the timing residuals of the model's prediction with the reported
data. The RMS of the reported residuals and the differences are $0.0086$ and $0.0048$, respectively, i.e., nearly a factor of two reduction of timing noise in terms of RMS with the application of the three-component model. In order to show the effectiveness of the three-component model, we perform and F-test for the three-component model and the base model adopted in the TEMPO2 program (Hobbs et al. 2006). The $F$ statistic is given by
\begin{equation}\label{f_test}
F=\frac{(\chi_1^2-\chi_2^2)/(d_1-d_2)}{\chi_2^2/d_2}\sim 27,
\end{equation}
where $\chi_1^2$ and $\chi_2^2$ are Pearson $\chi^2$ values, i.e. $\chi^2=\sum\frac{R_{i}^2}{\sigma_{i}^2}$, where $R_{i}$ is the residual of the $i$-th point, and $d_1=133$ and $d_2=124$ are the number of degrees of freedom for the base model and three-component model, respectively. Here we assume $\sigma_{i}=\sigma_{0}$, i.e, all data points have the same weight; this way, the result of the F-test is independent of the exact value of $\sigma_{0}$. $F\sim 27$ means that the probability to reject the three component model over the base model is less than $2.7\times 10^{-25}$, and thus the significance of the three-component model over the base model is higher than $10\sigma$. Our model implies that the timing residuals are also caused by the magnetic field oscillation, and the quasi-periodic structures in timing residuals have the same
origin (which is determined by Equation (\ref{braking law2})) as
those in $\ddot\nu$, $\dot\nu$ and $\nu$ variations. On the other hand, the fit is worse as the time span increases up to $\sim 30$ years, which may be mainly due to the additional noise components not included in $\ddot\nu$ variations.

The model includes an oscillation component
with a period of $\sim 49$ years, however, it is hard to test directly from the power
spectrum of its timing residuals, since the period is longer than
the observational data span. However, there are still some features
demonstrating its existence. For instance, the observed data were
reported about four years ago, and the model predicts
that $\ddot\nu$ of the pulsar is now experiencing another switch
from positive to negative (as shown in Figure \ref{Fig:1}), which can be
tested with the latest observed data. The test could also be
conducted by applying the model to a larger set of pulsars, which
have short oscillation periods (shorter than the observed time
span), and relatively large oscillation amplitudes (so that the
swing behavior of $\ddot\nu$ could emerge; the exact criteria of $k$
depend on $\nu$, $\dot\nu$ and $T$).

From Equation (\ref{braking law2}), we can obtain an analytic approximation from the one oscillation component model (in Paper I) for $\ddot\nu$
\begin{equation}\label{apprddot}
\ddot\nu\simeq -2\dot\nu(\alpha/t+ f \cos(\frac{2\pi t}{T}+\phi)),
\end{equation}
where $f=2\pi k/T$ represents the magnitude of the oscillation term. Thus, both parameters $k$ and $T$ are important. For $\alpha=0$, Equation (\ref{apprddot}) can be simply rewritten as
\begin{equation}\label{apprddot1}
\ddot\nu\simeq -2\dot\nu f \cos(\frac{2\pi t}{T}+\phi),
\end{equation}
One can see that the model predicts an oscillation behavior of $\ddot\nu$, which implies that one may get either a positive or a negative $\ddot\nu$.

From Table \ref{Tab:1}, we know that $f_1\approx f_3\gg f_2$ for PSR0329+54. Therefore the second component is less important in contributing to $\ddot\nu$, according to Equation~(\ref{apprddot1}). It is possible that the third component is the higher harmonic of the first component, since $T_1\approx 2T_3$. Thus it is likely that the first oscillation component dominates the timing behavior of the pulsar. As a matter of fact, there is almost always one dominant peak in the power spectrum of the timing residuals of most radio pulsars (H2010), i.e., one dominant oscillation component associated with their magnetic evolution.

A major prediction of this model with three oscillation components is that the averaged $\ddot\nu$ will start to decrease rapidly with additional data that extend just a few years beyond the span that was used in H2010, as shown by the black crosses in the upper panel of Figure~1. As the data are already available to the observers, we suggest that this prediction can be used to confirm or deny our model.

\section{Simulating the Distributions of $\ddot\nu$ and $n$ and their Correlations with $\tau_{\rm c}$}

In this section, based on our phenomenological model, we use the Monte Carlo method to simulate the distributions of $\ddot\nu$ and $n$, and
their correlation with $\tau_{\rm c}$. The ``averaging" effects are naturally included in the simulations. For simplicity, in the following simulation we assume that there is one dominant oscillation component, which mainly determines the variations of $\ddot\nu$ and the timing residuals, as discussed above. If the one oscillation component model is rejected by the reported data, then a multiple component model shall be presented. This is not in conflict with the above three components fit, since fitting to the distributions requires much less detailed information about variations in $\ddot\nu$ for individual pulsars.

We assume that the sample of phases $\phi$ of the field oscillation follows a uniform random distribution in the range from $-\pi$ to $\pi$. Randomly Drawing a data set $\{\nu, \dot\nu, T_{\rm s}\}$ from the reported sample space, i.e. from table 1 of H2010, calculating a corresponding start time $t_0$, and assuming some certain values for $k$ and $T$, we can obtain a rotation phase set $\{\Phi(t_i)\}$ using Equation (\ref{phase integrate}). In the calculation, the time interval for TOAs is also assumed to be a constant, i.e. $\Delta T_{\rm int}=10^6~{\rm s}$. Then the ``averaged'' values of $\nu$, $\dot\nu$ and $\ddot\nu$ can be obtained by fitting $\{\Phi(t_i)\}$ to Equation (\ref{phase}). Hence one has its averaged $|\ddot\nu|$, $|n|$ and $\tau_{\rm c}$. Repeat this procedure $N$ times, we will have $N$ data points in $|\ddot\nu|$-$\tau_{\rm c}$ and $|n|$-$\tau_{\rm c}$ diagram.

\subsection{Effects of Oscillation Period and Amplitude}

Analysis of a large sample of pulsar timing noise (H2010) showed that the oscillation periods are usually on the order of about $10$ years. However, the structures seem to vary with data span, and as more data are collected, more quasi-periodic features are observed. In this subsection, we investigate the ranges of variation of $k$ for a series of $T$.

\begin{figure}
\centering
\includegraphics[angle=0,scale=0.38]{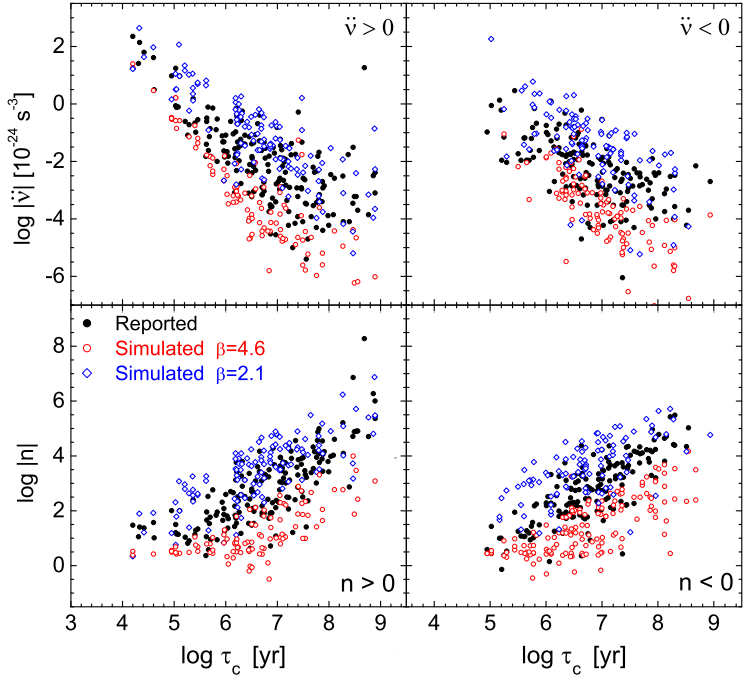}
\caption{Simulations of the $|\ddot\nu|$-$\tau_{\rm c}$ distribution (upper panels) and $|n|$-$\tau_c$ distribution (bottom panels) for $T=10$ years. } \label{Fig:4}
\end{figure}

We show the measured $|\ddot\nu|$ and $|n|$ versus $\tau_{\rm c}$ for $341$ normal radio pulsars with $\tau_{\rm c}<10^9~{\rm yr}$ in Figure {\ref{Fig:4}}, in which 184 pulsars with positive $\ddot\nu$ and $n$ are plotted in the left panels, and the other 157 pulsars with negative values in the right panels. The simulated results, for the case of $T=10$ yr, with $\beta=4.6$ or $2.1$ are also shown in the panels, in which $\beta$ is defined by $k=10^{-\beta}$ for convenience. One can see that the envelopes of $\beta=4.6$ and $2.1$ lie around the lower and upper boundaries of the reported data, respectively. This gives a natural constraint for $k$. Meanwhile, in the simulation the number of data points with $\ddot\nu>0$ should be roughly equal to the number of $\ddot\nu<0$, i.e. $N_{\rm p}/N_{\rm n}\simeq 1$. In Table \ref{Tab:2} we summarize the ranges of variation of $\beta$ for different $T$. Physically, $\beta<0$ is unacceptable, thus it should be $T\lesssim10^5$ yr. However, our model fails to give a tight constraint on $T$. Note that $\tau_{\rm c}$ in the figure is the characteristic age of the pulsars. However, $T_{\rm s}$ in Figure {\ref{Fig:1}} is the time span of the observation. Thus, the positive correlation between $\tau_{\rm c}$ and $n$ in Figure {\ref{Fig:4}} is not in conflict with Figure {\ref{Fig:1}} which shows $n$ decaying with $T_{\rm s}$.

\begin{table}[!h]
\tabcolsep 0pt \caption{Ranges of variation of $\beta$ for different $T$. $\beta_{\rm min}$ and $\beta_{\rm max}$ are the minimum and maximum values of $\beta$, respectively.} \vspace*{-12pt}
\begin{center}
\def\temptablewidth{0.8\textwidth}
{\rule{\temptablewidth}{1.1pt}}
\begin{tabular*}{\temptablewidth}{@{\extracolsep{\fill}}ccccccc}
 $T$ (years)                    & $10$      & $100$     & $1000$     & $10^4$     & $10^5$  \\
\hline
 ($\beta_{\rm min}$, $\beta_{\rm max}$) & (2.1, 4.6)& (1.9, 4.5)& (0.8, 3.4) & (0.3, 2.3) & (-0.5, 1.6)\\
       \end{tabular*}
       \label{Tab:2}
       {\rule{\temptablewidth}{1.1pt}}
       \end{center}
       \end{table}

\subsection{Two-dimensional Kolmogorov-Smirnov Test}

In this subsection, we perform a two-dimensional Kolmogorov-Smirnov (2DKS) test to reexamine the consistency of the distributions of the reported data and the simulated data, using the KS2D package\footnote{http://www.astro.washington.edu/users/yoachim/code.php}. If the returned p-value is greater than 0.2, then it is a sign that we can treat them as drawn from the same distribution.

\begin{figure}
\centering
\includegraphics[angle=0,scale=0.3]{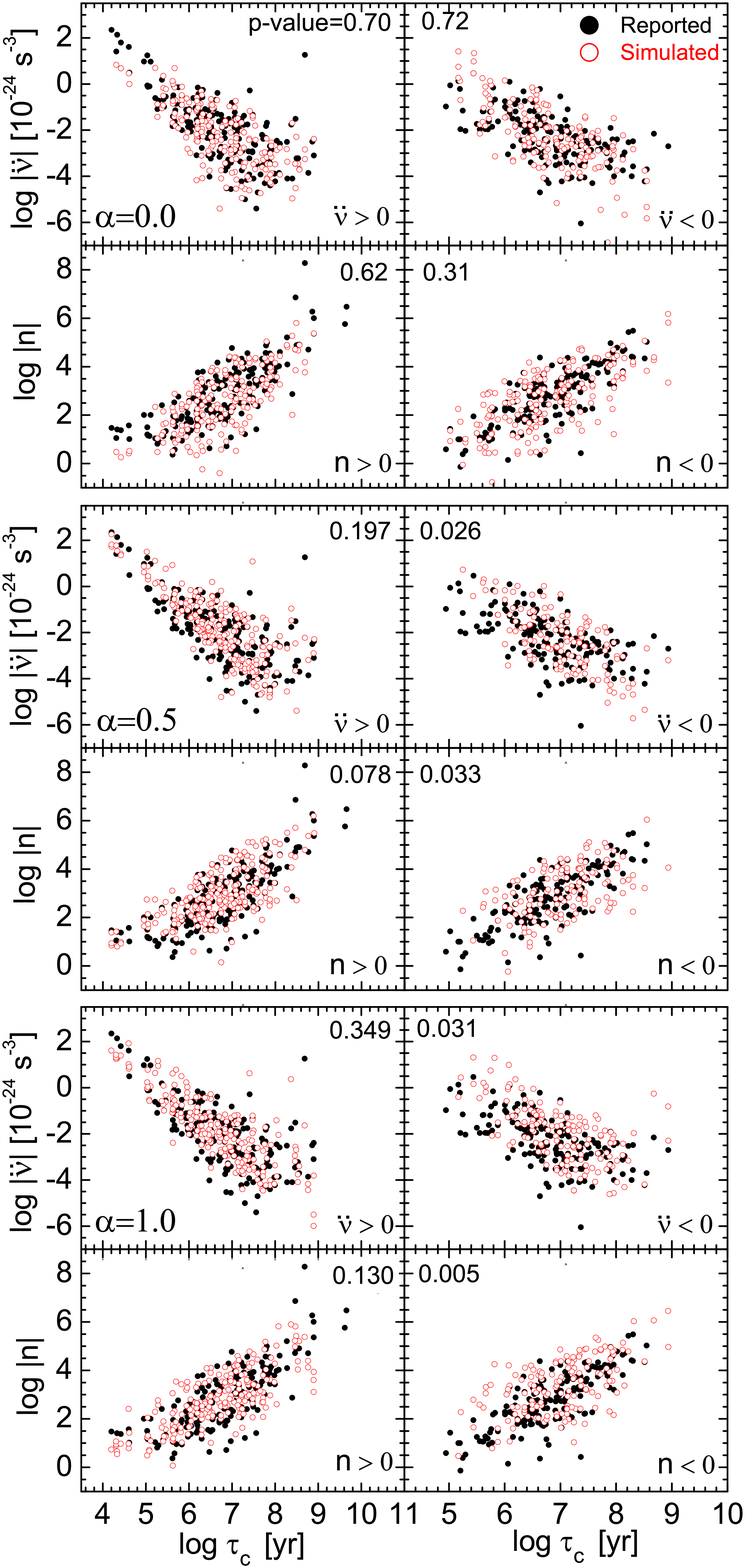}
\caption{Simulations of $|\ddot\nu|$-$\tau_{\rm c}$ and $|n|$-$\tau_c$ distributions for $T$ varying randomly from $1$ to $100$ years. The cases of $\alpha=0$, $\alpha=0.5$ and $\alpha=1$ are shown in the upper four panels, middle four panels and bottom four panels, respectively. } \label{Fig:5}
\end{figure}

We regard $T$ as a random number and allow it to vary from $1$ to $100$ years to account for the diversity of periodicities observed in the population. Let $\beta_{\rm min}$ and $\beta_{\rm max}$ vary from $1.5$ to $2.5$ and from $4.0$ to $5.0$, respectively. It is found that $\beta$ varying from $2.1$ to $4.5$ gives the highest p-value, as shown in the upper four panels of Figure \ref{Fig:5}. The returned probabilities are also labeled in each panel. Since the p-values indicate that the two samples are highly consistent, we thus conclude that the one oscillation component model with $\alpha=0$ is good enough to reproduce the $|\ddot\nu|$-$\tau_{\rm c}$ and $|n|$-$\tau_c$ distributions.

The simulated distributions with different $\alpha$ are also examined with a 2DKS test. We show the simulated distributions with $\alpha=0.5$ and $1.0$ in the middle and bottom four panels of Figure \ref{Fig:5}, respectively. To describe the evolution of the pulsar magnetic field, three routes are generally proposed (see e.g Goldreich \& Reisenegger 1992), i.e. ohmic dissipation, Hall effect, and ambipolar diffusion. Power law decay with $\alpha=0.5$ and $1.0$ are produced by ambipolar diffusion and Hall effect, respectively (Paper I). Here we do not include ohmic dissipation since it is not important for pulsars with $\tau_{\rm}\gtrsim 10^4 (\frac{B}{10^{12}~\rm G})^{-3}$~yr (Cumming et al. 2004). For both cases of $\alpha=0.5$ and $1.0$, the p-values of the simulated data for $\ddot\nu<0$ are much lower than $0.2$, and thus are rejected by the test. In fact, one can see that for $\tau_{\rm c}\gtrsim 10^6$~yr there is a crowded area of data points along the lower boundary for $\ddot\nu>0$, and the data points are scarce around the lower boundary for $\ddot\nu<0$. This is mainly caused by the long-term magnetic field decay, i.e. the decay term $-2\dot\nu\alpha/t>0$ dominates the oscillation term $-2\dot\nu f\cos(\frac{2\pi t}{T}+\phi)$ in Equation \ref{apprddot} for some cases. However, there is no such crowded area or scarce area in the reported data, which clearly indicates that the model with $\alpha=0$ is preferred for pulsars with $\tau_{\rm c}\gtrsim 10^6$~yr.

\section{Summary and Discussion}

In this work we first modeled the $\ddot\nu$ and $n$ evolutions and
applied the obtained model parameters to simulate the timing
residuals for the individual pulsar PSR B0329+54. Using a Monte
Carlo simulation method, we simulated the distributions of pulsars
in the $|\ddot\nu|-\tau_{\rm c}$ and $|n|-\tau_{\rm c}$ diagrams,
and compared the simulation results with the reported data in H2010.
Our main results are summarized as follows:

\begin{enumerate}

\item We modeled the $\ddot\nu$ evolution of pulsar PSR B0329+54
with a phenomenological model that incorporates the evolution of $B$, which
contains three oscillation components (upper panels of Figure~\ref{Fig:1}). The
model can reproduce the $|\ddot\nu|$ variation quite well, including the swings between $\ddot\nu>0$ and
$\ddot\nu<0$. This model predicts that the averaged $\ddot\nu$ of PSR B0329+54 will start to decrease rapidly with newer data beyond those used in H2010.

\item We showed that the ``averaged'' values of $n$ are different from the
instantaneous values (bottom panels of Figure~\ref{Fig:1}), and the
oscillation abruptly decays after the first period due to the
``averaging'' effect. Using these parameters obtained from modeling
the evolution of the averaged $\ddot\nu$, we simulated the timing residuals of the
pulsar (Figure~\ref{Fig:3}), which agrees with the reported
residuals (H2010) well.

\item We performed Monte Carlo simulations for the distribution of
$|\ddot\nu|$ and $|n|$ in the $|\ddot\nu|-\tau_{\rm c}$ and $|n|-\tau_{\rm c}$ diagrams, respectively. The simulated results
for different modes of long-term decay of the magnetic field (i.e.
$\alpha=0$, $\alpha=0.5$ and $1.0$ in Figure~\ref{Fig:5}) are tested by the 2DKS. It is found that the reported distributions can be well reproduced with the one-oscillation-component model with $\alpha=0$ for pulsars with $\tau_{\rm c}\gtrsim 10^6$~yr.

\end{enumerate}

Pons et al. (2012) proposed a similar model of magnetic field oscillations with a timescale of $(10^6-10^8)\frac{10^{12}~{\rm G}}{B}~{\rm yr}$ and magnitude $\delta B/B\sim10^{-3}$, and obtained pulsar evolutionary tracks in the $P-\dot P$ diagram. Lyne et al. (2010) showed credible evidence that timing residuals and $\dot\nu$ are connected with changes in the pulse width. Therefore, timing residuals are more likely caused by changes in a pulsar's magnetosphere with periods of about $1-100~{\rm yr}$. In Xie \& Zhang (2013), we suggested that perturbations from Hall waves in the dipole magnetic field associated with NS crusts are probably responsible for the observed quasi-periodic oscillations in the timing data as well as changes in the pulse width, which may provide a physical explanation for the present model.

We therefore conclude that magnetic field oscillations dominate the long term spin-down behaviors of old NSs, for which the long-term field decay is not important, in contrast to younger NSs with $\tau_{\rm c}\lesssim 10^6$~yr. The fact that only one oscillation component is required to reproduce the observed $|\ddot\nu|-\tau_{\rm c}$ and $|n|-\tau_{\rm c}$ distributions suggests that there is one dominant oscillation component for most NSs, and thus does not conflict with the fact the multiple oscillation components are also often observed in some pulsars. In fact, for some pulsars, the structures seen in the timing noise vary with data span and more quasi-periodic features are observed for longer data span (H2010). Admittedly, our current model cannot predict the number, amplitudes and periods of oscillation modes. However, our model can adequately describe the acquired timing data with a small number of oscillation modes, as shown in section 2, which represents a first step towards understanding of the magnetic field oscillations of NSs. As such, our understanding of the oscillation modes will be improved as more quasi-periodic features are revealed with longer observations in the future. In addition, our model can also describe the distributions of $\ddot\nu$ and $n$ reasonably well. As far as we are aware of, our work is the first one in which the distribution of $\ddot\nu$ is used to test the long-term magnetic field evolution of NSs, which is independent from tests based on the traditional $\nu-\dot\nu$ diagram.

\normalem
\begin{acknowledgements}
We thank Shuxu Yi and Meng Yu for valuable discussions. SNZ acknowledges partial funding support by the National Basic Research Program of China (973 program, 2009CB824800), by the National Natural Science Foundation of China under grant Nos. 11133002, 11373036 and 10725313, and by the Qianren start-up grant 292012312D1117210.

\end{acknowledgements}

\end{document}